# A User Study to Investigate Semantically Relevant Contextual Information of WWW Images[1]

Wan Fariza Fauzi & Mohammed Belkhatir

Faculty of IT, Monash University

wfpai1@student.monash.edu; belkhatir.mohammed@infotech.monash.edu

**ABSTRACT**

The contextual information of Web images is investigated to address the issue of enriching their index characterizations with semantic descriptors and therefore bridge the semantic gap (*i.e.* the gap between the low-level content-based description of images and their semantic interpretation). Although we are highly motivated by the availability of rich knowledge on the Web and the relative success achieved by commercial search engines in indexing images using surrounding text-based information in webpages, we are aware that the unpredictable quality of the surrounding text is a major limiting factor. In order to improve its quality, we highlight contextual information which is relevant for the semantic characterization of Web images and study its statistical properties in terms of its location and nature considering a classification into five semantic concept classes: signal, object, scene, abstract and relational. A user study is conducted to validate the results. The results suggest that there are several locations that consistently contain relevant textual information with respect to the image. The importance of each location is influenced by the type of webpage as the results show the different distribution of relevant contextual information across the locations for different webpage types. The frequently found semantic concept classes are object and abstract. Another important outcome of the user study shows that a webpage is not an atomic unit and can be further partitioned into smaller segments. Segments containing images are of interest and termed as image segments. We observe that users typically single out textual information which they consider relevant to the image from the textual information bounded within the image segment. Hence, our second contribution is a DOM Tree-based webpage segmentation algorithm to automatically partition webpages into image segments. We use the resultant human-labeled dataset to validate the effectiveness of our segmentation method and experiments demonstrate that our method achieves better results compared to an existing segmentation algorithm.

## 1. INTRODUCTION

In the recent years, there has been a dramatic increase in the number of images due to advance storage technology as well as affordable digital cameras. Furthermore, we have the World Wide Web (WWW) making these images accessible on a global scale. The WWW has transformed the way people communicate and socialize; the new wave of social networking websites such as Facebook, MySpace, Flickr, YouTube etc. are enabling people to upload and share images and other multimedia contents on the WWW. Thus, the WWW can be viewed as an inexhaustible collection of images across diverse domains. These developments have heightened the need to come up with better image retrieval systems to manage and locate this heterogeneous collection of images. In this article, we consider the problem of highligting information relevant to the image content on the WWW for indexing and retrieval purposes.

Generally, there are two approaches of image retrieval systems: content-based and concept-based. Content-based image retrieval (CBIR) systems are non-linguistics based. They do not require images to be indexed with textual labels, instead images are indexed and retrieved based on their low-level visual features of colors, shapes, texture, and/or spatial information [11, 14, 36]. While CBIR systems are fully automatic and fast, they pose a problem as humans perceive images at a higher level, in terms of semantic concepts, therefore making CBIR suitable only for domain-specific applications. The difference in

---

interpretation is commonly known as the semantic gap [35]. Another disadvantage of CBIR is the querying method, since images are indexed with low-level features, users can only query by sketch, color composition and/or example (*i.e.* by providing an example image). The query by sketch method is not desirable for non-artistically inclined users and the query by example method assumes that the user has an example on hand. Words are still predominant in image indexing and retrieval systems, at least in the foreseeable future [31].

The second approach, called concept-based, relies on linguistics and involves the abstract identification of high-level semantic concepts in an image that are expressed with textual descriptions, such as "beach", "city", "tigers", "death of a pop star" etc. Traditional keyword-based image retrieval systems, CBIR systems with high-level semantics and web-based image retrieval systems could be categorized under concept-based image retrieval systems.

The traditional keyword-based systems involve the manual tagging of images with keywords. This task, initially costly and impractical as the corpus size increases, is now achievable with the Web 2.0 technology, as implemented in Flickr, the online photo sharing website. The main disadvantage of this method is that it is highly dependent on the indexer; the annotations are error-prone, not comprehensive and the range of successful queries is limited to the interpretation of the indexer [18, 35].

CBIR systems with high-level semantics are essentially CBIR systems that are able to learn high-level semantic concepts from low-level visual features using computer vision and machine learning techniques, focusing on reducing the semantic gap. Such systems include semantic-based image retrieval systems [5, 21, 23] and signal/semantic-based systems [3, 29]. Bradshaw *et al.* [5] use a probabilistic model in their semantic-based system to recognize four concepts (*i.e.* natural/man-made and indoor/outdoor) in an image. Jeon *et al.* [21] implement a cross-media relevance model and identify 70 object concepts. Li *et al.* [23] recognize 101 object categories. The drawback of these systems is the loss of visual information in the learning process. Signal/semantic-based systems address this loss by retaining the relation between object concepts and the visual information; Belkhatir *et al.* [3] identify 72 semantic concepts, 11 color categories and 11 high-level textures. The major limiting factor of such systems is that they can only recognize a fixed set of semantic classes. In Liu *et al.*'s survey [26] on CBIR with high-level semantics, they mention Li *et al.*'s 101 object categories [23] form the largest vocabulary set used in object recognition. This figure is far from the 5,000 – 30,000 object categories perceived by humans [26]. Clearly, these systems are still a distance away from closing the semantic gap.

We consider web-based image retrieval systems as concept-based systems since web images are typically indexed with textual descriptions in view of the fact that they come with rich contextual information, which is their associated surrounding text, used jointly with their filename, alt description, and page title. For instance, current commercial WWW image search engines such as Google Image, Alta Vista and Yahoo! use this contextual information to index and retrieve images. Though these systems have achieved real world success, their retrieval precision is highly dependent on the quality of the image contextual information and can sometimes be poor (*c.f.* figure 1), as a result, users have to go through pages of search results to find the desired image.

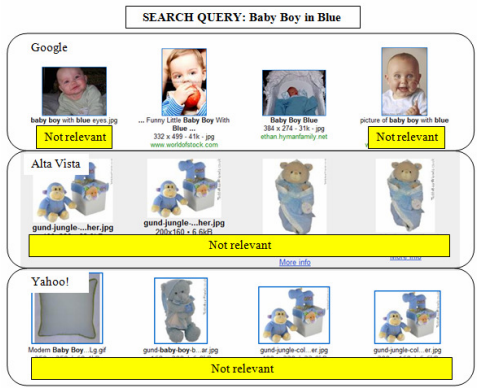

**Figure 1. The first rows of retrieved results from three current WWW image search engines**

There are apparent differences in various categories of image retrieval systems, with each having its own strengths and weaknesses. As such, many researchers are exploiting on the strengths of each category through fusion and multimodality, consequently, addressing one or the other system's limitations.

Back to our focus on the WWW and its assorted image collection, there are many research works that involve the fusion of visual features from CBIR systems and the contextual information of web images [6, 10, 13, 25, 27, 33, 38]; they draw on the contextual information as a source of countless semantic concepts to increase the cardinality of the sets of semantic classes.

They attempt to address at the same time the poor retrieval performance of current web-based systems with the image visual content.

While Cai *et al.* [6], Gao *et al.* [13] and Wang *et al.* [40] combine textual and visual information to cluster images, the works of Sclaroff *et al.* [33], Lu *et al.* [27], Feng *et al.* [10], Liu *et al.* [25] and Wang *et al.*[38] are concerned with the fusion of the two sources for image annotation purposes. In the earlier papers [27, 33], the textual and visual information are kept in separate data stores and loosely coupled by means of relevance feedback where the user is required to feedback which images are relevant (or irrelevant) out of all the images in his/her search result, making such systems less attractive due to unnatural interaction. Feng *et al.* [10] and Liu *et al.* [25] strongly couple both textual and visual image contents using a bootstrapping approach and graph learning method, respectively. They improve the system recall rate and effectively remove the need for user feedback. However, there is only a minor increase in the system retrieval precision. A possible reason for this is the inaccuracy of the textual information attached to the image, which brings us back to the original problem of unpredictable quality of the contextual information of a Web image faced by the current web-based systems.

This makes the need to understand the contextual information of a Web image crucial. A review on the literature reveals two main problems affecting the quality of the image contextual information. Firstly, there is no standard definition. While the image filename, alt description and page title are generally accepted by researchers in the area, the difference is in the surrounding text, which has been perceived as a window of words [19, 32], a paragraph [12, 34], a section/segment [6, 10, 20, 24] and even the entire page [13, 16]. The differences of opinions amongst researchers can be attributed to the fact that webpage partitioning is a difficult problem due to the diverse page layouts. Secondly, there is a lack of semantic techniques in filtering out irrelevant words from the contextual information. Standard text processing techniques are commonly employed to filter out irrelevant words [10, 13, 17, 20, 25]. These techniques include stemming, stop-word removal, retention of noun phrases, frequent words and/or co-occurring words. Thus, we require more knowledge of the characteristics of the relevant image contextual information.

We perform two tasks: (i) a webpage observation exercise and (ii) a user study, with the aim to elucidate the definition of the contextual information of a web image and present a clear analysis on the characteristics of relevant contextual information in terms of its location within a HTML source code and its semantic nature according to a classification framework of image descriptors, therefore addressing the problems above. The initial webpage observation is performed by the authors of this paper on 386 images from webpages randomly picked from 5 website categories – Business, Informational, News, Advocacy and Personal [37]. We establish the image surrounding information as a section containing the image and a variable number of words. Within a section, words relevant to each image are identified and analyzed. We present some general characteristics of the data. These characteristics include distributions of locations of the relevant contextual image information and the classification of relevant contextual image information into five semantic concept classes: signal, object, scene, abstract and relational, with their distributions. The subjectivity in associating the relevant words to an image compels us to conduct a user study to evaluate our work. Hence, a user study, involving 33 subjects, is carried out on another dataset of 898 images, also randomly selected from the five categories of webpages. Again, the distribution tables for the location and semantic nature of relevant contextual information are presented. The results from the user study undoubtedly supersede the initial analysis, but the initial analysis has dictated the methodology of the user study and which aspects of the problem are to be considered in the user study.

The user study also supports the assumption that the webpage can be partitioned into smaller sections and addresses the problem of ambiguity in defining the boundary of the contextual information for web images. At this point, we present a DOM Tree-based webpage segmentation algorithm that automatically segments webpages into sections, with each section consisting of a web image and its contextual information (*i.e.* image segment). The state-of-the-art segmentation techniques are discussed in detail in the next section. The proposed segmentation method can extract image segments from a diverse range of websites automatically without the need for training, thus making it practical and scalable. Experimental results indicate that our method outperforms an existing state of the art segmentation algorithm, VIPS [7], in precision and recall. While this segmentation method can solve the inadequate usage of contextual text by web search engines to annotate images illustrated in Figure 2, it is still crucial to deal with the noise (*i.e.* irrelevant words) present in the extracted contextual information. In the future, we anticipate addressing this with the prior probabilities of the location and semantic nature of the relevant image contextual information presented in this paper, with the hope of improving the quality of the image contextual information. This information, in turn, can be applied directly to annotate images whereby we envisage the classification of contextual information into the 5 semantic concept classes to contribute to a multi-facetted image retrieval model, giving users the flexibility to choose which facets of an image they wish to query (e.g. querying for an abstract concept of the image). This information can also be used as a source of textual information in a multi-modal image retrieval system, enriching semantic characterization from the image visual features.

## 2. RELATED WORK

Highly regarded works in WWW image retrieval systems acknowledge that WWW images have rich contextual information in their hosting webpages and many systems use combined models to support Web image retrieval.

In earlier works, Sclaroff *et al.*'s ImageRover [33] use textual and visual cues. The textual cues come from image filename, ALT attribute within the  tag of the HTML file, link string, and title of the HTML page. ImageRover includes the surrounding text, which is defined as the 10 words appearing before the  tag and 20 words appearing after the  tag. Emphasized words in bold and italics, word frequency and word proximity to the image are all taken into account using the Latent Semantic Indexing (LSI) method. For word proximity, only 10 and 20 words appearing before and after the image respectively are considered. Relevance feedback is used to further improve the search results.

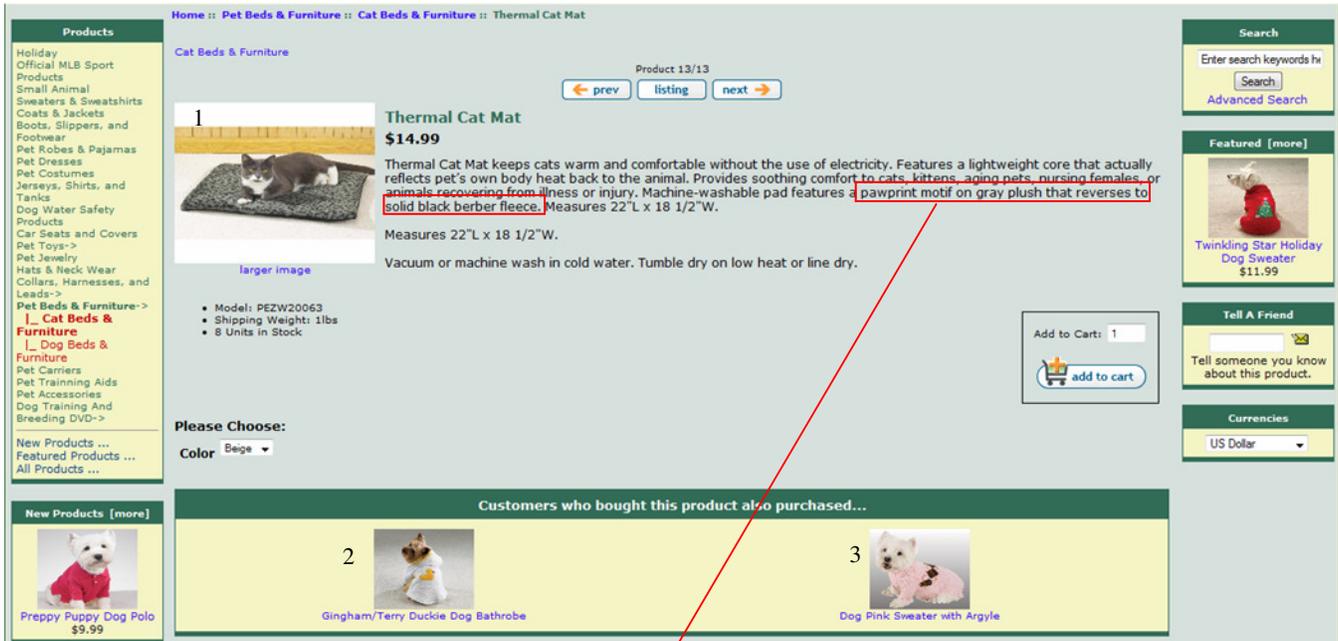

(a) A random webpage with URL: www.spoilurpets.com/thermal-cat-mat-p-150.html

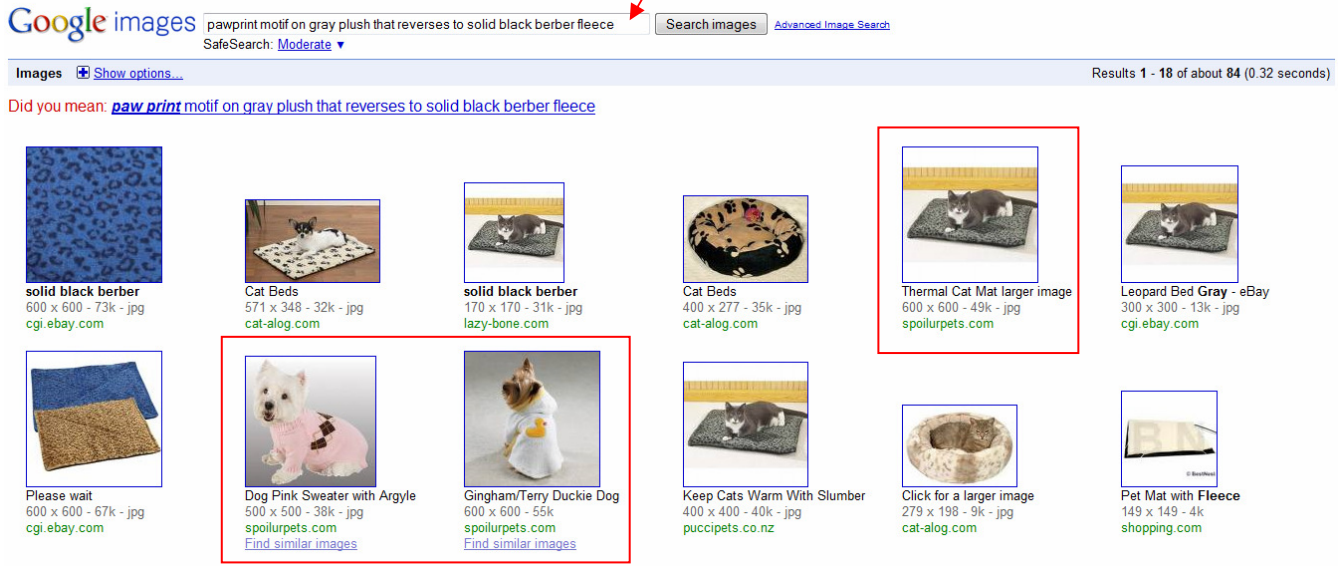

(b) First page of the search result for the query "pawprint motif on gray plush that reverses to solid black berber fleece"

**Figure 2. An example of the usage of contextual text by Google Image to annotate image. The surrounding texts highlighted in the red box in (a) are used as the query in (b) and we can see image 2 and 3 are wrongly indexed with those texts.**

Shen *et al.* [34] use image filename, image ALT, page title and image caption. The latter is defined as the entire paragraph containing the image. They represent the semantics of the image using the surrounding texts and relevance feedback mechanisms. The surrounding texts are weighted using lexical chains in a Weight ChainNet model. Chen *et al.* [8] consider the entire paragraph as the source of contextual information similarly to Shen *et al.* [34]. In this work, the textual representation is combined with low-level visual features of the image.

Feng *et al.* [10] carry out a study and define the surrounding texts as a section of texts separated by structural HTML tags such as <TABLE>, <TR>, <TD>, <DIV> and <HR> with a cutoff point at a text description length greater than 32 words before and after an image. While efficient, the heuristics used above work on limited webpages, and fall back on fixed window size. The texts are then fused together with visual features using a co-training approach. Similarly, Hua *et al.* [20] also define surrounding texts as a section of texts but rely on the border properties of structural HTML tags as separators instead. Scalability and consistency of these systems require further investigation.

In Cai *et al.* [6], the VIPS (Vision-based Page Segmentation) algorithm is implemented to partition a webpage according to its visual presentation into smaller blocks/sections. All the textual information within a section containing an image is defined as the image contextual information. Each image is represented using text, visual and link information. This information is used to cluster or organize the search result but does not address the problem of using textual information to annotate images.

A couple of recent works in Gong *et al.* [15] and Gao *et al.* [13] consider the entire page as a source for extracting the image contextual information. Gong *et al.* [15] index images based solely on the surrounding texts classified into 3 categories – TM (texts from the title and meta tags), LT (texts attached to the image tag) and BT (texts of the body which are big in size). Texts from BT are segmented in blocks according to tags considered to be semantic delimiters (<body>, <table>, <p>, <ul>, <div>, etc.) and weights are attached to each block. Gao *et al.* [13] focus on image clustering based on visual features and surrounding texts. The surrounding texts include all texts from the entire webpage excluding the stop words. This system manipulates certain categories of webpages which are the photography museums and galleries only and thus might not be scalable to other types of websites.

Amidst the differences in definition of contextual image information, it is clear that no standard work exists as how to associate text to WWW images. However, we can observe that common locations of contextual image information are image SRC, image ALT, page title and nearby texts that range from 20 words to the entire page.

To extract the image contextual information, generally there are two methods. The first and simplest method is to use a fixed window size of minimum 20 terms to the entire page [13, 15, 33]. The second method performs webpage segmentation to extract sections containing the images and their contextual information [6, 10, 20, 24]. There are two approaches to webpage segmentation: i) DOM Tree-based and ii) DOM Tree-based with additional visual information obtained from rendering the DOM Tree. Typically, the webpage DOM tree structure is analyzed to discover section-specific patterns. Both Feng *et al.* [10] and Hua *et al.* [20] implement the DOM Tree-based method. The existing DOM Tree-based methods require further improvement to accommodate diverse webpages. Cai *et al.*'s VIPS algorithm [6] is an example of a DOM Tree-based approach incorporating additional visual information. Li *et al.* [24] also include visual cues in their page segmentation algorithm. Even though visual cues might improve accuracy, these algorithms are known to be computationally expensive and become crucial when processing the large-scale Web.

Some existing Web information extraction (IE) systems are reviewed. There has been considerable research on the general problem of extracting information from webpages. The latter are considered as semi-structured pages. Most Web IE systems look into automating the translation of the semi-structured input webpages into structured data, ready for post processing. They rely on the fact that the majority of dynamic webpages have some forms of underlying templates. Works in [2,9] focus on the automatic extraction of structured data from webpages relying on automatically-generated templates [2] or wrappers [9].

In our context of extracting WWW images and their contextual information, images can be found in many types of websites, not just data-intensive websites, but in unique webpages as well. Images that are inconsistent with the layout of the input HTML pages will be lost in the data extraction process.

To date, in progressing towards formulating a standard definition of the contextual information relevant for the content of Web images, which comes in many varieties in terms of page layout and style, no one has yet taken into account the WWW users' viewpoints. All related works involving user studies were mainly to evaluate completed image retrieval systems and assess their retrieval performance with respect to the needs of its users. In view of the differences of opinions in the related work, we suggest that the users are brought into the loop to provide their perception on this issue.

# 3. ANALYSIS OF IMAGE CONTEXTUAL INFORMATION

An initial investigation on image contextual information was conducted by the authors to study its characteristics. With this knowledge, some hypotheses were formed. This initial investigation consisted of an observation exercise on 50 webpages. Specific questions that are addressed in the observation exercise:

1. What is the contextual information for a web image?
2. How many words to consider as the contextual information?
3. Are all the words that are considered as contextual information relevant to the web image?
4. What are the characteristics of the relevant contextual information?

## 3.1 Dataset

Ideally, we would like to observe all images from all webpages as our target is the WWW in general (*i.e.* to be able to extract any web image and its contextual information from any webpage). However, considering the number of webpages available on the WWW, this is impossible. Hence, to ensure a wide coverage of webpages, we examine Tate et al.'s Web taxonomy [37]. The Web is classified into five categories from a Library Science perspective, where webpages are considered as a crucial library resource. The five types of webpages defined are: Business, Informational, News, Advocacy and Personal. This categorization was made in the nineties. Since then, many other types of websites have materialized that could still be categorized in any of the 5 types. For example, in the Alexa Web Directory (http://www.alexa.com/) which uses DMOZ, the most comprehensive human-reviewed directory of the web, webpages under the *shopping* category are considered as Business pages; those under *movies*, *health*, *education*, *government* as Informational Pages and *weblogs* as Personal pages. For the purpose of covering many types of webpages, this course-grained taxonomy is sufficient.

386 images of reasonable dimension (images with both width and height being at least 50 pixels) are collected from 50 webpages randomly picked from the abovementioned categories. Embedded dynamic images are excluded and beyond the scope of this project.

## 3.2 Method

Putting ourselves in a Web user's position, we scrutinized the webpages. We were inclined towards the definition of image contextual information as a section of texts in [6, 10, 20] rather than a fixed number of words before and after an image or words in the entire webpage. We also observed two distinct patterns of Web images embedded within webpages – unlisted (*c.f.* Figure 3 – image 1) and listed images (*c.f.* Figure 3 – image 2, 3, and 4). *Unlisted images* are standalone or random images that appear anywhere on a webpage and *Listed images* are two or more images that are systematically ordered within the webpage.

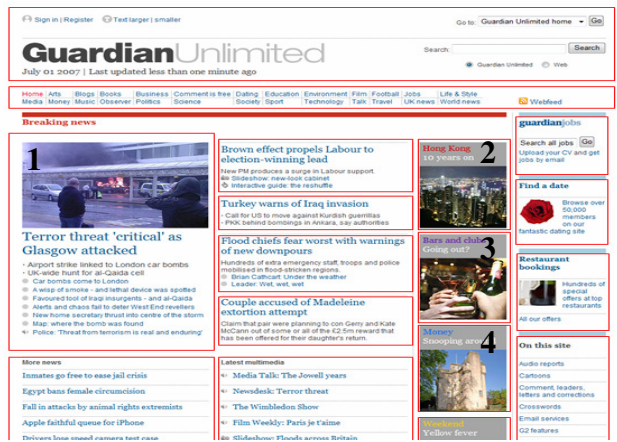

**Figure 3. Example page partitioned to smaller units.**

Based on these observations, we manually partitioned each webpage into smaller sections which we thought were semantically related (*c.f.* Figure 3). Sections with images are of importance to us and these sections are termed as *image segments*. Within each image segment, we further investigated the contextual information to answer the third question – are all the words that are considered as contextual information (*i.e.* words within each segment) relevant to the web image?

Our investigation showed that only a subset of the contextual information was relevant to the images as highlighted in the rectangular boxes in Figure 4. Next, we examined the corresponding HTML source code for each image segment to obtain the locations of the identified relevant words in terms of HTML tag as well as to consider the HTML tag attributes which include the common locations of contextual information mentioned in the related works: (1) image SRC, (2) image ALT, and (3) page title. The HTML tag attributes are typically hidden from the Web user and viewable only in the HTML source code.

A quantification study was carried out to evaluate the probability of appearance of the relevant surrounding words in those locations and to determine the nature of the relevant surrounding words in terms of semantics at word or phrase level.

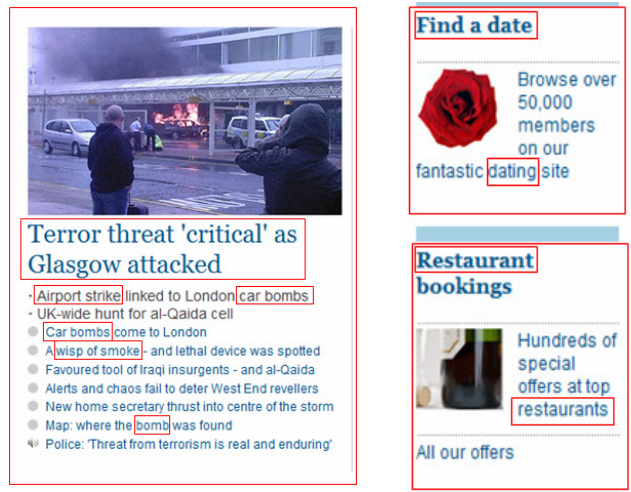

**Figure 4. Some examples of WWW images and their semantically relevant contextual information.**

### 3.2.1 Locations of Relevant Contextual Image Information

From the examination of the HTML source code, we found relevant words in the SRC and ALT attributes of  as expected and many more relevant words in other tags as well such as the anchor tag <A>. We discovered relevant words that were located further than 20 words, 32 words, and even further than a paragraph (*c.f.* Figure 5). Hence, using a fixed window size to correlate nearby texts to images is not applicable to all WWW images; a dynamic webpage segmentation algorithm is required.

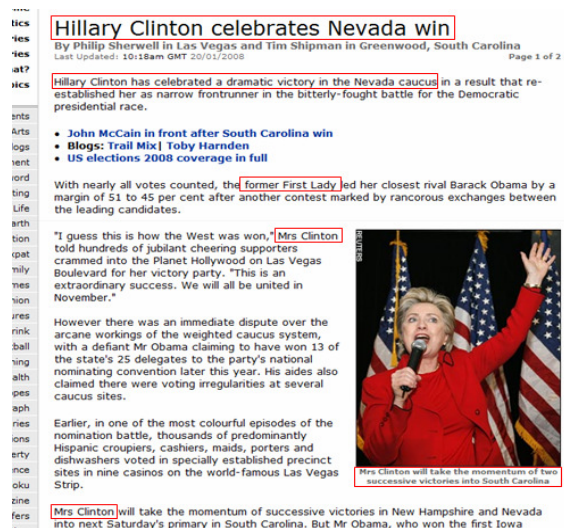

**Figure 5. Example of semantically relevant words found outside the paragraph containing the image.**

The frequency distributions of all the locations where we can find contextual information relevant to the WWW images are constructed and organized according to the five webpage categories of Business, Informational, News, Advocacy and Personal. Table 1 shows the frequency distribution tables for Business and Personal webpages. The set of locations of relevant contextual image information varies across the five webpage categories.

The locations of relevant contextual information can be categorized into two types: visible and hidden. Visible surrounding words are those that could be viewed on the webpage via a Web browser. Typically, these are *strings enclosed by* the following HTML tags: <TITLE>, <BODY>, <P>, <A>, <DIV>, <TD>, <LI>, <DT>, <DD>, and <LABEL>. Hidden surrounding words are found in the attributes of the HTML tags viewable only through the source file of the webpage, for example, ALT, SRC, NAME of an  tag, HREF, TITLE, NAME of an <A> tag, etc.

Under each category in Table 1, the locations that contain substantial relevant contextual information are highlighted in yellow. Our aim is to design an efficient algorithm so that time is not spent unnecessarily on extracting and processing texts from locations that have a low likelihood of containing relevant contextual image information. As such, from the distribution tables, it can be hypothesized that for each webpage category, only the highlighted locations are specially significant and meaningful for our extraction module to extract relevant contextual image information.

**Table 1. Distribution of relevant texts based on location for web images from five categories**

|  | Business | Info | News | Advocacy | Personal |
|---|---|---|---|---|---|
| **ALT of ** | 26.7 | 12.1 | 11.7 | 26.4 | 7.3 |
| **SRC of ** | 21.5 | 11.4 | 11.7 | 27.6 | 6.9 |
| **TITLE of ** | 0.2 | 0 | 0.2 | 0.0 | 0.9 |
| **CLASS of ** | 0 | 0.1 | 0.1 | 0.0 | 0 |
| **LONGDESC of ** | 0 | 0.7 | 0 | 0.0 | 0 |
| **ID of ** | 0 | 0 | 0.2 | 0.0 | 0 |
| **NAME of ** | 0 | 0 | 0 | 1.1 | 0.2 |
| **ALT of <AREA>** | 1.1 | 0 | 0 | 0.0 | 0 |
| **HREF of <AREA>** | 1 | 0 | 1.4 | 0.0 | 0 |
| **HREF of <A>** | 16.8 | 38.8 | 17.6 | 5.7 | 22.5 |
| **ONCLICK of <A>** | 0.5 | 0 | 7.5 | 0.0 | 4.2 |
| **TITLE of <A>** | 0.9 | 0.6 | 0.1 | 2.3 | 9.7 |
| **OBJECTID of <A>** | 0.1 | 0 | 0 | 0.0 | 0 |
| **ALT of <A>** | 0 | 0 | 0.9 | 0.0 | 0 |
| **NAME of <A>** | 0 | 0 | 12.1 | 0.0 | 0 |
| **ONMOUSEOVER of <A>** | 0 | 0 | 0 | 0.0 | 0.5 |
| **ONMOUSEOUT of <A>** | 0 | 0 | 0 | 0.0 | 0.2 |
| **CLASS of <DIV>** | 0.4 | 0.3 | 0.2 | 0.0 | 0 |
| **TITLE of <DIV>** | 0.2 | 0 | 0 | 0.0 | 0 |
| **ID of <DIV>** | 0 | 0.3 | 0.4 | 0.0 | 0.2 |
| **ID of <INPUT>** | 0 | 0 | 0.2 | 0.0 | 0 |
| **VALUE of <INPUT>** | 0 | 0 | 0.2 | 0.0 | 0 |
| **SRC of <INPUT>** | 0 | 0 | 0.1 | 0.0 | 0 |
| **NAME of <INPUT>** | 0 | 0.1 | 0.1 | 0.0 | 0 |
| **VALUE of <OPTION>** | 0 | 0.7 | 0 | 0.0 | 0 |
| **CLASS of <LI>** | 0 | 0 | 0.5 | 0.0 | 0 |
| **ID of <H2>** | 0 | 0 | 0.1 | 0.0 | 0 |
| **FOR of <LABEL>** | 0 | 0 | 0.1 | 0.0 | 0 |
| **ACTION of <FORM>** | 0 | 0 | 0.2 | 1.1 | 0 |
| **string enclosed by <A>** | 15.1 | 18.3 | 22.8 | 10.3 | 12.2 |
| **string enclosed by <TD>** | 9.5 | 1.3 | 0.2 | 18.4 | 0.4 |
| **string enclosed by <DIV>** | 3.4 | 4.9 | 3.1 | 2.3 | 6.6 |
| **string enclosed by <BODY>** | 0 | 0 | 0 | 0.0 | 0.9 |
| **string enclosed by <P>** | 0.2 | 8.8 | 4 | 2.3 | 15.2 |
| **string enclosed by <LI>** | 0.1 | 0 | 2.8 | 0.0 | 0.2 |
| **string enclosed by <OPTION>** | 0 | 0.4 | 0 | 0.0 | 0.2 |
| **string enclosed by <BLOCKQUOTE>** | 0 | 0 | 0 | 0.0 | 1.5 |
| **string enclosed by <DT>** | 0 | 0 | 0.2 | 0.0 | 0 |
| **string enclosed by <DD>** | 0 | 0 | 0.2 | 0.0 | 0 |
| **string enclosed by <LABEL>** | 0 | 0.1 | 0.3 | 0.0 | 0 |
| **<SCRIPT>** | 1.8 | 0.1 | 0.8 | 0.0 | 1.5 |
| **Comment** | 0 | 0 | 0 | 0.0 | 6.3 |
| **CONTENT of <META>** | 0.4 | 0.4 | 0 | 0.0 | 0.4 |
| **string enclosed by <TITLE>** | 0.1 | 0.1 | 0 | 2.3 | 1.8 |
| **TITLE of <LINK>** | 0 | 0 | 0 | 0.0 | 0.2 |
|  | **100** | **100** | **100** | **100** | **100** |

Interestingly, we found several locations that consistently contain relevant contextual information irrespective of the page category. These locations are ALT and SRC of , HREF and TITLE of <A>, and strings enclosed by <A>, <TD>, <DIV> and <P> (shown in the blue rows in Table 1). This is particularly useful for extracting contextual image information from webpages that do not fall in any of the five categories. We now turn to the analysis on the nature of the semantically relevant contextual information.

*3.2.2 Nature of Relevant Contextual Image Information*

The semantically relevant contextual image information found can be a single word, a phrase or a sentence. For extracted sentences, a linguistic parser is used to highlight noun and verb phrases. An image concept is defined as a word or phrase expressing the image content. An image is itself described by one or more concepts.

Jaimes and Chang.[1] proposed a ten-level image descriptor model consisting of: type/technique, color, texture, shape, generic-object, generic-scene, specific-object, specific-scene, abstract-object, abstract-scene. Hollink *et al.* [18] described 3 levels of image descriptors: non-visual, perceptual and concept level. The concept level is further divided into 3 sub-levels: general, specific and abstract.

In our work, we consider a five-concept-type characterization. The five concept types are: signal, object, scene, abstract and relational.

- Signal concepts are concepts that can be extracted automatically. These concepts are usually the low-level visual features such as color, shape, texture and spatial words (e.g. pink striped or whirly blue).
- Object concepts are words that describe the individual entities in the images; objects could be of living (e.g. human, animals, plants, etc.) or non-living entities (e.g. table, fan, book, etc.).
- Scene concepts are words that describe the image as a whole based on the set of all of the objects it contains [1] (e.g. city, landscape, indoor, outdoor, still life, portrait, mountain scene, beach, field etc).
- Abstract concepts are intangible concepts that represent knowledge or additional information including abstract characteristics (e.g. emotions, quality) regarding the image or image entity. Abstract concepts also include symbolic concepts that do not visually describe the image but are deemed relevant (e.g. the word "travel" is considered relevant for an image containing the objects bag and hat).
- Relational concepts are concepts that associate two or more objects in terms of visible action or non-visible information about the image or image entity such as author, brand, photographer, date and type of image, etc.

We provide in Figure 6 a few examples of images with their characterization in terms of semantic concepts. Table 2 presents the classification of relevant contextual information into the five concept types.

**Table 2. Concept type distribution of the semantically relevant contextual information**

| *Category* | *Business* | *Info* | *News* | *Advocacy* | *Personal* |
|---|---|---|---|---|---|
| **Signal** | 2.3 | 0.4 | 0.3 | 1.1 | 0.7 |
| **Object** | 70.8 | 49.9 | 49.0 | 62.1 | 75.2 |
| **Scene** | 0.1 | 10.8 | 12.2 | 5.7 | 3.5 |
| **Abstract** | 7.0 | 35.8 | 32.2 | 28.7 | 14.5 |
| **Relational** | 19.8 | 3.0 | 6.3 | 2.3 | 6.1 |
| **Total (%)** | 100.0 | 100.0 | 100.0 | 100.0 | 100.0 |

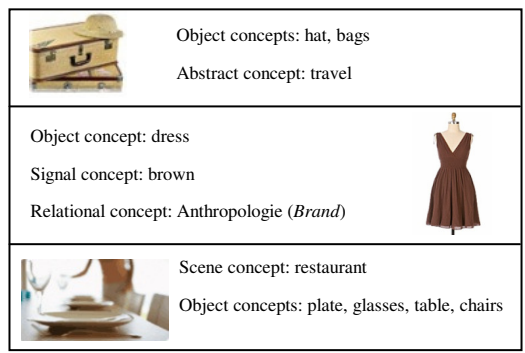

**Figure 6. Semantic Concept characterization of some example images**

Knowing the distribution of concepts will give us an idea on how to characterize the semantic nature of the extracted contextual information of WWW images and quantify its importance.

It is apparent from Table 2 that very few signal concepts, which are used in the framework of content-based indexing, are found within the relevant contextual image information. Object concepts are dominant, followed by abstract concepts.

Considering an existing image indexing framework mapping low-level visual features to high-level general concepts (such as people, animal, plant...), a plausible approach is to use the available Wordnet ontology to integrate the general concepts to more specific concepts using the hypernymy semantic relation. This is most suited for object and scene concepts. Relational concepts are used to provide relationship between object concepts. Abstract concepts are the hardest to include. We observed that they are usually the text links that link to the same webpage as the image link. Another observation is that the abstract concepts usually appear near the related object/scene concepts. For example, an image of Bill Clinton giving a speech has the relevant contextual information: *Clinton hits rival on foreign affair*. Here, *Clinton* (a noun phrase) is the object concept followed by the abstract concept: *hits rival on foreign affair*. Further data analysis is required in the future to associate abstract concepts to images.

In term of image indexing, the concept distribution is essential for selecting which concepts are to be used. Hence, from the distribution tables, it can be hypothesized that for each webpage category only the highlighted concepts are specially significant and meaningful.

A user study is conducted to validate our findings which will be explained next.

## 4. USER STUDY / VALIDATION

### 4.1 Aim of Study

In the previous section, a web observation exercise was performed to define the characteristics of the relevant contextual information. Two hypotheses were formed:

H1: On the location of contextual image information for each webpage category, only the highlighted locations in Table 1 are specially significant and meaningful for our extraction module to extract relevant contextual image information.

- H1-1: For business pages, the important locations are: ALT and SRC of , HREF of <A>, strings enclosed by <A> and <TD>.
- H1-2: For information pages, the important locations are: ALT and SRC of , HREF of <A>, strings enclosed by <A> and <P>.
- H1-3: For news pages, the important locations are: ALT and SRC of , HREF and NAME of <A> and strings enclosed by <A>.
- H1-4: For advocacy pages, the important locations are: ALT and SRC of , strings enclosed by <A> and <TD>.
- H1-5: For personal pages, the important locations are: HREF and TITLE of <A>, strings enclosed by <A> and <P>.

H2: On the nature of contextual image information for each webpage category, only the highlighted concepts in Table 2 are specially significant and meaningful to be included in an image index.

- H2-1: For business pages, the important concepts are: object and relational concepts.
- H2-2 & H2-3: For information and news pages, the important concepts are object, scene and abstract concepts.
- H2-4 & H2-5: For advocacy and personal pages, the important concepts are object and abstract concepts.

We conducted a user survey to gather the users' perception on the semantically relevant contextual information of WWW images. The outcome will support or reject the above hypotheses.

We used a binomial test to validate our shortlisted list of important locations and semantic nature.

### 4.2 Methodology

An exploratory-cum-descriptive user study was conducted to investigate user perceptions of WWW images and their contextual information. The user study consisted of a two-part questionnaire: the first part is a set of questions to ease the subject towards the tasks ahead as well as to gather the subject's experience in searching images. This part identifies the subject's level of familiarity with the WWW and experiences with current image search engines.

The second part is to evaluate the assumptions and hypotheses stated in section 4.1. The subject is presented with screenshots of ten webpages, which include all the five types of webpages. For each webpage, the subject must perform three tasks – a "segmentation" task (if the subject agrees that a webpage consists of multiple topics), an "extraction" task and a "describe" task.

In the first task, the subject studies a webpage and decides whether the webpage is a single-topic block or is made up of multiple single-topic blocks. If he/she thinks it contains multiple blocks, then he/she is required to circle all the blocks. From this first task, we can verify whether the assumption that a webpage can be segmented into smaller blocks is supported.

The second task is related to the hypotheses on the location and nature of the contextual information and gives some evidence to support our assumption that small and advertisement images are noise images. For this task, the subject identifies all the images in the webpage and for each image, selects words or phrases which he/she thinks is relevant to the image. Additionally, he/she needs to provide a brief description of the image, which is actually the third and final task of the user study.

The "describe" task addresses the fourth question where we ask the subject to give a free text description of an image. This "describe" task allows us to highlight any relevant words from the hidden contextual information when there is a match between the subject's image description and the hidden surrounding words.

A pilot user study was carried out to evaluate how well the subject understood the questions in the survey and whether they were able to complete the tasks as well as the amount of time required to complete them. The pilot user survey was carried out on three subjects. The results confirmed that the subjects were able to understand the questions and complete the tasks well. The amount of time taken to complete the task was around 60 to 90 minutes.

The actual user study was then conducted on 40 subjects and a total of 100 webpages were investigated. With each subject scrutinizing 10 webpages, we had 4 subjects performing the same task on each webpage. At the end, we have 4 sets of observations for 100 webpages.

The recruited subjects were students and lecturers from local universities as well as their family members and friends. Only 33 subjects of the 40 recruited subjects completed the survey. Thirteen males and twenty females, aged between 18 and 52, were involved. All of them were familiar with the Internet and previously used it to search for images. Furthermore, 63% of the subjects thought that the current image search engines were partially effective.

### 4.3 Data Analysis and Discussion
All images and their associated relevant contextual information plus the users' free descriptions were tabulated.

#### 4.3.1 Segmentation
In the first task where the subjects were asked whether each webpage was a single-topic block or was made up of multiple single-topic blocks, we found that most information and advocacy pages were considered single-topic blocks because the subjects felt that these pages provided many aspects of a single general topic (e.g. medical information, movie information, education page, parenting, and cancer-related pages). However, they still partitioned the webpage into header, navigation, content, and footer blocks and when they picked out images and contextual information, none of the subjects considered the entire page in associating word/phrase/sentences to the image. All the subjects chose words/phrases/sentences that were near the images. Figure 7 shows an example of a user's perception of images and their semantically relevant contextual information. As such, we can safely assume that a webpage can indeed be partitioned into smaller semantic units.

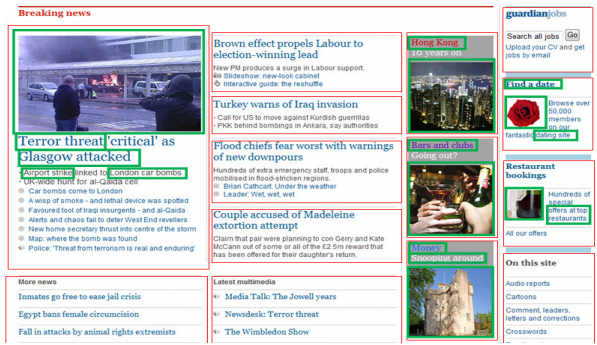

**Figure 7. Example of a subject's perception of images and their semantically relevant contextual information**

*4.3.2 Images and Their Semantically Relevant Contextual Information*

From the second task where the subjects were asked to pick out images and semantically relevant contextual information, a total of 898 images were identified with relevant contextual information; out of which only 550 images were used for data analysis. The images identified only by a single subject were left out of the analysis. Images that were mostly ignored by the subjects were small images, advertisement images, and text images. Only 18% of the subjects extracted very small images and 9% included advertisement images. The remaining 82% of the subjects identified images with width and height greater than 45 pixels and having a paragraph close to them. Other images that were excluded were images embedded in flash and JavaScript objects. These embedded objects are beyond the scope of our current research.

For all images in the dataset, each image had 0-4 surrounding words or phrases (surrounding descriptors) selected as relevant to the image. Each image was also described with a few words or phrases by the subjects (subjects' descriptors). The descriptors were then counter-checked with the respective webpages to ensure they were legitimate/valid surrounding words, thus reducing errors. Surrounding descriptors that could not be found in the HTML source code were excluded from the list whereas subjects' descriptors found near the image were included in the surrounding descriptor list. Then, each of the descriptors was associated to its location within the HTML source of the webpage.

For each category of webpages, the frequency distribution of relevant surrounding descriptors for each location was calculated. The contextual information of an image was modeled as a set, denoted by S. By default S is an empty set, S = { }. If a subject identifies a contextual word/phrase semantically relevant to the image content, then set S has an element, S = {$s_1$}. If another subject picks out two words/phrases one of which is identical to $s_1$, then one new element is added to the set S, S = {$s_1$, $s_2$}. After all new elements were added to the set, each element was associated to its number of occurrences and respective location(s) within the HTML source code. This was repeated for all images in all webpages per category.

**Table 3. Distribution of relevant texts based on location found in 5 categories of webpages from the user study**

| | Business | Info | News | Advocacy | Personal |
|---|---|---|---|---|---|
| **ALT of ** | 17.9 | 10.2 | 12.6 | 11.3 | 3 |
| **SRC of ** | 19.4 | 10.7 | 7.9 | 9.4 | 6.8 |
| **TITLE of ** | 0.3 | 0.3 | 0.6 | 0.6 | 0.3 |
| **ID of ** | 0 | 0.2 | 0 | 0 | 0 |
| **NAME of ** | 0 | 0 | 0 | 0.6 | 0 |
| **ONMOUSEOVER of ** | 0 | 0 | 0 | 0.4 | 0 |
| **ALT of <AREA>** | 0.1 | 0.6 | 0 | 0 | 0 |
| **HREF of <A>** | 15.4 | 27.9 | 11.5 | 16.5 | 31.1 |
| **TITLE of <A>** | 0.2 | 5.9 | 0.7 | 1.5 | 7 |
| **ONCLICK of <A>** | 0 | 0 | 3.4 | 1.3 | 3.5 |
| **ALT of <A>** | 0 | 0 | 1.2 | 0 | 0 |
| **NAME of <A>** | 0.1 | 0 | 11.8 | 0 | 0 |
| **CLASS of <DIV>** | 0.1 | 0 | 0 | 0 | 0 |
| **TITLE of <DIV>** | 0.2 | 0 | 0 | 0 | 0 |
| **ID of <DIV>** | 0 | 0.2 | 0 | 0 | 0 |
| **CN of <DIV>** | 0 | 0 | 0.6 | 0 | 0 |
| **VALUE of <INPUT>** | 0 | 0 | 0.1 | 0 | 0 |
| **ID of <UL>** | 0.1 | 0 | 0 | 0 | 0 |
| **CLASS of <LI>** | 0.6 | 0 | 0.2 | 0 | 0 |
| **ID of <LI>** | 0.1 | 0 | 0 | 0 | 0 |
| **ACTION of <FORM>** | 0.1 | 0 | 0 | 0 | 0 |
| strings enclosed by <A> | 19.8 | 20.2 | 32.2 | 9.2 | 14.3 |
| strings enclosed by <TD> | 12.2 | 5.2 | 0.9 | 8.1 | 1.3 |
| strings enclosed by <TH> | 0 | 0.1 | 0 | 0 | 0 |
| strings enclosed by <DIV> | 6.2 | 5.3 | 9.1 | 9.6 | 8.3 |
| strings enclosed by <CITE> | 0 | 0.3 | 0 | 0 | 0 |
| string enclosed by <BODY> | 0 | 0 | 0 | 0 | 1.6 |
| strings enclosed by <P> | 1.1 | 8.5 | 3 | 14.6 | 17.8 |
| strings enclosed by <LI> | 1.5 | 0 | 1.3 | 3.8 | 0.3 |
| string enclosed by <OPTION> | 0 | 0 | 0 | 0.4 | 0.1 |
| string enclosed by <DT> | 0 | 0 | 0.5 | 0 | 0 |
| string enclosed by <DD> | 0 | 0 | 0.4 | 0 | 0.3 |
| **<SCRIPT> tag** | 0 | 0.1 | 0.1 | 1.2 | 1 |
| comment tag | 0.3 | 0 | 0.1 | 1.9 | 0.2 |
| **CONTENT of <META>** | 3.3 | 3.6 | 1.2 | 6.3 | 0.9 |
| strings enclosed by <TITLE> | 1 | 0.7 | 0.6 | 3.3 | 2.2 |
| | **100.00** | **100.00** | **100.00** | **100.00** | **100.00** |

The frequency distribution for each webpage category was computed and tabulated in Table 3. Next, the relevant contextual information identified by the users was classified into the five semantic concept types and presented in Table 4.

Table 4. Concept distribution of the relevant contextual information from the user study

| Category | Business | Info | News | Advocacy | Personal |
|---|---|---|---|---|---|
| Signal | 2.5 | 0.4 | 0.2 | 0 | 1.9 |
| Object | 69.6 | 46.6 | 39.8 | 44.9 | 55.8 |
| Scene | 0 | 7.9 | 10.1 | 4.3 | 5.1 |
| Abstract | 23.4 | 39.1 | 41.3 | 47.6 | 26.0 |
| Relational | 4.5 | 6.0 | 8.6 | 3.2 | 11.2 |
| Total (%) | 100.0 | 100.0 | 100.0 | 100.0 | 100.0 |

Before further analysis was carried out to validate our hypotheses, we verified that the size of the user and data sets were large enough using the "Split-half Analysis Consistency" [28]. The split-half analysis consistency test is one of the many types of reliability tests. It randomly splits the collected data into 2 sets, X and Y, and computes the reliability coefficient, $r$ between the two sets. A reliability coefficient of 0.70 or higher is considered acceptable reliability in statistical research. If $r > 0.70$, then we can conclude that the sample size is sufficient to give reliable results.

For 550 Web images, we randomly split the data into two sets and computed the reliability coefficient for each category – Business ($r = 0.84$), Information ($r = 0.91$), News ($r = 0.88$), Advocacy ($r = 0.9$), and Personal ($r = 0.92$), which shows that the data obtained under each webpage category are sufficient.

For 33 subjects, we obtained at least 2 sets of observations for 550 Web images. We computed the Pearson's correlation value for the 2 sets, which is equal to 0.87, indicating that the data obtained is highly consistent as well. Thus, there is no need for replacement of the 7 non-respondents.

### 4.3.3 Statistical Validation Using Binomial Test

In section 4.1, we stated that for each page category, there are several locations more important and meaningful than the rest because they contain more semantically relevant contextual image information (H1 hypothesis). We performed the Binomial Test on the user survey data to validate our hypotheses.

A Binomial Test is an exact test of statistical significance of deviation of a sample proportion from a theoretically-expected proportion. The condition of using the binomial test is that the sample must consist of dichotomous data – the individual (i.e. the concept) in the sample is classified in one of two categories (i.e. location under test/the rest of the locations). This test is used to evaluate/discover if the proportion of individuals falling in each category differs from chance or from some pre-specified probabilities of falling into those categories [4].

For us to use the binomial test, all the locations identified in Table 1 were tested independently. Thus, the concepts in the user survey data were either located in the location under test (LUT) or in the rest of the other locations (OTHERS).

Using the "equal opportunity for each location" hypothesis, we established the null and alternate hypotheses.

$H_0$: *There is no reason to consider that the LUT is especially significant for our extraction algorithm, which means that all the relevant contextual information is equally distributed between all possible locations.*

$H_A$: *The LUT is significant for our extraction algorithm.*

The 5% significance level ($\alpha = 0.05$) was used to accept or reject the null hypothesis. We will explain in the following section the Binomial Test procedure and results.

**Binomial Test Procedure**
For business pages (refer to Table 1), of the 19 locations of relevant contextual image information we identified in our Web observation exercise, we hypothesized that the meaningful and significant locations containing most of the relevant contextual information are ALT of , SRC of , HREF of <A>, string enclosed by <A> and <TD>.

From the users' standpoint, we want to validate the observation. The first location under test is the ALT of  (LUT = ALT of ). Since we are trying to prove that the location ALT of  is one of the important locations, our null and alternate hypotheses are:

$H_0$: There is no reason to consider that the ALT of  is especially significant for our extraction algorithm, which means that all the relevant contextual information is equally distributed between all possible locations (i.e. each with a probability of $\frac{1}{19} = 0.053$).

$H_A$: The ALT of  is significant for our extraction algorithm.

The null and alternate hypotheses provided symbolically are:

$H_0 : p = p(ALT) <= 0.053$ and $q = p(OTHERS) >= 0.947$

$H_A : p > 0.053$ (and $q < 0.947$)

We checked if the normal approximation can be used by calculating the proportion of individuals in the sample expected to fall into each category multiplied by the total number of individuals in both categories (represented by n), denoted by *np* and *nq*. Both *np* and *nq* must be greater than 10 (*i.e. np > 10* and *nq > 10*) to use the normal approximation.

Here, we have *n*=905, *p*=0.053 and *q*=0.947. So *np* = 47.965 and *nq* = 857.035.

Since both *pn* and *nq* are greater than 10, we can use the normal approximation to the binomial distribution. With α = 0.05, the critical region is defined as any z-score value greater than +1.645 considering that we are performing a one tailed hypothesis test.

The test statistic is computed as follows:

$$z = \frac{X/n - p}{\sqrt{pq/n}} = \frac{SampleProportion(Data) - HypothesizedProportion}{Std\_error}$$

where the *SampleProportion(Data)* is the proportion of relevant concepts in the user survey data that is located in the LUT which corresponds to probability values found in Table 3. For example, for LUT = ALT of , 162 relevant concepts out of 905 relevant concepts are located in ALT of , so the *SampleProportion(Data) is 162/905 =0.179*. Hence, the z-score is:

$$z = \frac{162/905 - 0.053}{\sqrt{(0.053)(0.947)/905}} = \frac{0.179 - 0.053}{0.0074468} = 16.92$$

The obtained z-score is in the critical region. Therefore, we reject the null hypothesis. On the basis of these data, we concluded that the percentage of relevant contextual image information in ALT of  is significantly higher than in any other locations if we were to have an equal distribution between all possible locations. Hence, the location ALT of  is evidently an important and meaningful location for the extraction of semantically relevant contextual image information as compared to the rest of the locations and the result is statistically significant at the *0.05* level of significance. We repeated the binomial test for other locations tested at α = 0.05; the result is shown in Table 5.

Table 5. Binomial Test results for locations found in BUSINESS pages

| Location | z-score | Conclusion |
| --- | --- | --- |
| ALT of  | 16.92 | Reject $H_0$ |
| SRC of  | 18.99 | Reject $H_0$ |
| TITLE of  | -6.67 | Accept $H_0$ |
| ALT of <AREA> | -6.97 | Accept $H_0$ |
| HREF of <AREA> | -7.12 | Accept $H_0$ |
| HREF of <A> | 13.51 | Reject $H_0$ |
| ONCLICK of <A> | -7.12 | Accept $H_0$ |
| TITLE of <A> | -6.82 | Accept $H_0$ |
| OBJECTID of <A> | -7.12 | Accept $H_0$ |
| CLASS of <DIV> | -6.97 | Accept $H_0$ |
| TITLE of <DIV> | -6.82 | Accept $H_0$ |
| string enclosed by <A> | 19.44 | Reject $H_0$ |
| string enclosed by <TD> | 9.20 | Reject $H_0$ |
| string enclosed by <DIV> | 1.19 | Accept $H_0$ |
| string enclosed by <P> | -5.63 | Accept $H_0$ |
| string enclosed by <LI> | -5.19 | Accept $H_0$ |
| <SCRIPT> tag | -7.11 | Accept $H_0$ |
| CONTENT of <META> | -2.67 | Accept $H_0$ |
| string enclosed by <TITLE> | -5.78 | Accept $H_0$ |

The Binomial Test procedure is repeated for the remaining four webpage categories. The results obtained are summarized in Table 6.

**Table 6. Summary of binomial test results on the location hypotheses of the semantically relevant contextual image information**

| Category | Hypotheses | User Survey |
|---|---|---|
| Business | H1-1:<br>ALT of <br>SRC of <br>HREF of <A><br>strings in <A><br>strings in <TD> | ALT of <br>SRC of <br>HREF of <A><br>strings in <A><br>strings in <TD> |
| Information | H1-2:<br>ALT of <br>SRC of <br>HREF of <A><br>strings in <A><br>strings in <P> | ALT of <br>SRC of <br>HREF of <A><br>strings in <A><br>strings in <P> |
| News | H1-3:<br>ALT of <br>SRC of <br>HREF of <A><br>NAME of <A><br>strings in <A> | ALT of <br>SRC of <br>HREF of <A><br>NAME of <A><br>strings in <A><br>strings in <DIV> |
| Advocacy | H1-4:<br>ALT of <br>SRC of <br>strings in <A><br>strings in <TD> | ALT of <br><br><br>HREF of <A><br>strings in <P> |
| Personal | H1-5:<br>HREF of <A><br>TITLE of <A><br>strings in <A><br>strings in <P> | HREF of <A><br>TITLE of <A><br>strings in <A><br>strings in <P><br>strings in <DIV><br>SRC of  |

For both business and information pages, the binomial test on the user survey data showed that the locations we have considered important for the extraction of semantically relevant contextual information are also considered important by the users. For news pages and personal pages, additional important locations are highlighted: strings enclosed by <DIV> for news pages; strings enclosed by <DIV> and SRC of  for personal pages. Only for advocacy pages, the only mutual location considered important is the ALT of . For all webpage categories, we will use the locations evaluated as important and meaningful from the user survey since the dataset used in the survey is larger than the dataset used in the Web observation exercise.

**Table 7. Summary of binomial test results on the concept hypotheses of the contextual image information**

| Category | Hypotheses | User Survey |
|---|---|---|
| Business | H2-1:<br>Object concepts<br>Relational concepts | Object concepts<br><br>Abstract concepts |
| Information | H2-2:<br>Object concepts<br>Scene concepts<br>Abstract concepts | Object concepts<br><br>Abstract concepts |
| News | H2-3:<br>Object concepts<br>Scene concepts<br>Abstract concepts | Object concepts<br><br>Abstract concepts |
| Advocacy | H2-4:<br>Object concepts<br>Abstract concepts | Object concepts<br>Abstract concepts |
| Personal | H2-5:<br>Object concepts<br>Abstract concepts | Object concepts<br>Abstract concepts |

As mentioned previously in section 3.2.1, we found several locations in Table 1 (shown in blue rows) that consistently contain semantically relevant contextual information irrespective of the webpage category. Following a quick comparison of Table 1 and Table 3, again we could observe the same pattern; these locations are present in all page types. Hence, for pages that cannot be classified within any of the five categories, the contextual image information will be extracted from ALT and SRC of , HREF and TITLE of <A>, and strings enclosed by <A>, <TD>, <DIV> and <P>.

Similarly, the binomial test is performed on the semantic concept distribution from the user survey.

Table 7 shows that users, without fail, picked object concepts to relate to images. Additional knowledge (abstract concepts) related to the objects in the image are equally important to the users for all categories of pages. Hence, an image indexing model shall include both object and abstract concepts to comply with the users' perception.

## 5. A FRAMEWORK FOR AUTOMATIC EXTRACTION OF WEB IMAGES AND THEIR CONTEXTUAL INFORMATION

In the previous sections, we presented our effort to highlight contextual information semantically relevant to the content of WWW images. We characterized this information in terms of location (*i.e.* where to locate it in the webpage) and nature (*i.e.* what type of semantic concepts we can find in the webpage). We conducted a user survey to validate our characterization.

Based on the findings of the user survey, we propose a novel framework for automatically extracting the contextual information for the content of Web images. The outcome of the "segmentation task" in Section 4.3.1 substantiates our assumption that a webpage can be partitioned into smaller segments; the image segments highlighted by the users are in line with our observations in Section 3.2, whereby two distinct types of Web images embedded within webpages were observed - unlisted and listed images. These user-identified image segments form the human-labeled dataset for testing our proposed segmentation algorithm.

Further investigation on the different types of Web images is required to formulate the automatic segmentation algorithm. Thus, the Document Object Model (DOM) of the webpage is considered. The webpage is parsed by a browser to obtain its DOM Tree structure. The DOM Tree is examined to discover the third type of Web images – semi-listed images, with each type having unique DOM Tree patterns. Hence, the three types of images are formally defined as follows:

- *Unlisted images* are standalone or random images that appear anywhere on a page (*c.f.* Figure 8a: Segment 1), e.g. profile photos in personal homepages, company logos, advertisements etc. The corresponding DOM Tree consists of an image node with its surrounding text as text node siblings and a root HTML tag representing the boundary of this image segment (*c.f.* Figure 8b).

- *Listed images* are two or more images that are systematically ordered within the webpage (*c.f.* Figure 8a: Segment 1-8). Examples of listed images are representative images, list of product images, news images, etc. The associated DOM Trees for such image segments are characteristically comprised of the image node and the nodes of the surrounding text forming a sub-tree structure under a root HTML tag. Other siblings under this root HTML tag share similar sub-tree structures (*c.f.* Figure 8d).

- *Semi-listed images* are visually similar to listed images. The difference is characterized by their DOM trees. DOM trees of semi-listed images are similar to DOM Trees of unlisted images in the sense that the image node and the nodes of the surrounding text are under a root HTML tag representing the segment boundary. But along with those nodes, there are other image nodes with surrounding text nodes at the same level (*c.f.* Figure 8c).

### 5.1 Algorithm

We propose a novel DOM Tree based segmentation algorithm to automatically extract image segments from webpages using the image characteristics mentioned above. For every image node found in the DOM Tree, the algorithm finds the image segment using a heuristic determined by the image characteristics. This is accomplished by detecting the variation in the total number of texts in each upward level of the DOM Tree, beginning from the image node. We use Segment 1 from Figure 8a to explain this. From the image node, the algorithm traverses up the DOM Tree, and stops at the \*<TABLE> node, where it first detects text nodes. An increase from zero to one text node can be seen from the respective DOM Tree structure in Figure 8d. The change denotes the first image segment, which is the solid rectangular box highlighting Segment 1. The algorithm continues upwards until it detects another change in the number of text nodes at the \*<TR> node where a bigger segment is detected as shown in the dashed box encompassing Segment 1 and 2 in Figure 8a. Sibling sub-trees are checked for listed and unlisted images. The smaller segment is regarded as the boundary of the relevant contextual information for listed images and the larger segment for unlisted images. Sub-trees containing listed images will have siblings with similar sub-tree structures.

In our example, the image is a listed image; therefore the smaller segment, which is the solid rectangular box, is taken as the segment boundary for segment 1.

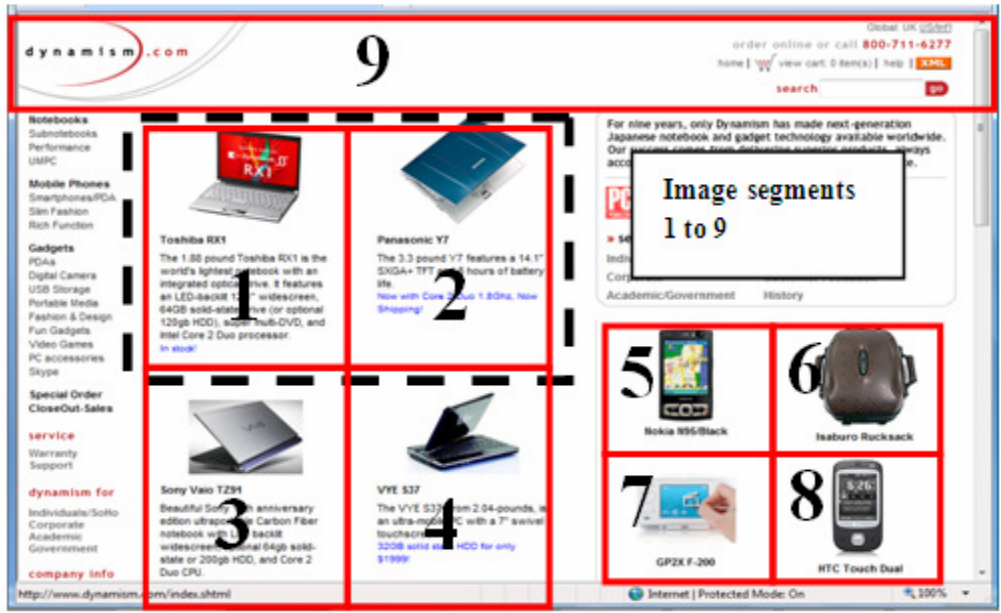

**(a) Image segments 1 - 9**

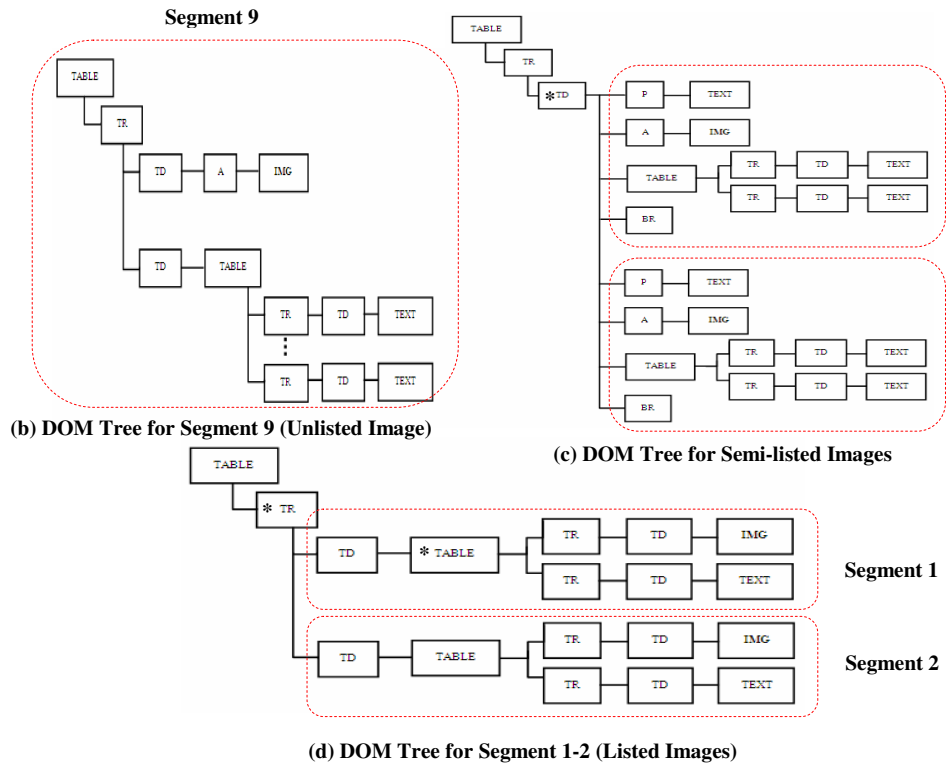

**(b) DOM Tree for Segment 9 (Unlisted Image)**

**(c) DOM Tree for Semi-listed Images**

**(d) DOM Tree for Segment 1-2 (Listed Images)**

**Figure 8. Example of image segments and their corresponding DOM Tree Structures**

The algorithm summarizing the automatic segmentation of webpages for the extraction of the semantically relevant image contextual information is as follows:

| | |
|---|---|
| **Algorithm:** The Segmentation Algorithm for Web Images ||
| **Require**: *I* ← The set of valid image nodes from a webpage. ||

```
 1: for all i_k ϵ I do
 2: repeat
 3:    int stateImg = getNumImage();
 4:    int stateText = getNumTextNode();
 5:    int state = 0;
 6:    if (stateText!=state && stateImg>0 && stateText>0)
 7:       if (stateChangeTwice)
 8:          if (parentOfListedImage)
 9:             take childNode as region
10:          else (parentOfUnlistedImage)
11:             take parentNode as region
12:          end if
13:       else
14:          if (SectionsInSameLevelForSemiListedImage)
15:             partition sections accordingly
16:          else
17:             state = stateText;
18:             stateChangeTwice = true;
19:          end if
20:       end if
21:       childNode = parentNode;
22:       parentNode = parentNode.getParentNode();
23:    end if
24: until parentNode != null
25: end for
```

The algorithm is iterative. It starts with the image node to identify image segments and stops when all valid images are processed. An image segment must contain at least one image node and one text node. The *stateImg* and *stateText* variables keep track of the number of image and text nodes respectively, while *state* variable records the current state. Our algorithm detects segments, by comparing the *stateText* variable to the *state* variable. The initial segment and its bigger super-segment are detected by the *stateChangeTwice* variable. The image characteristics influence the decision on the choice of the finalized image segment.

Upon detection of the initial segment, our algorithm checks for semi-listed images whose sections occur in the same level of the DOM Tree (*c.f.* Figure 8c). We search for patterns and separation points that divide the semi-listed images and their contextual information. If these sections exist, we proceed to partition them into individual sections and extract them accordingly. For example, in Figure 8c, the initial section is detected at node *<TD>, we check for repeating patterns in this sub-tree and here, the nodes repeat themselves in the sequence of <P>, <A>, <TABLE> and <BR>. Upon identifying the sequence, the separation points are determined to partition them according to regions. In this example, the starting cutoff point is set to node <P> and an ending cutoff point is set to node <BR>. Consequently, the semi-listed images are partitioned into two sections shown in the dashed boxes. Otherwise, the algorithm will resume upward traversal until another change in the number of text nodes is detected, indicating that a bigger section is found. At this point (*i.e. stateChangeTwice* is true), we check for listed and unlisted images. If listed images are found, we take the smaller section as the image segment; otherwise, we take the bigger section as our choice.

Therefore, based on the variations found in the number of text nodes at different DOM Tree levels, our algorithm detects image segments and extracts them accordingly.

## 5.2 Experiments and Results

Contextual image information has long been mined for various uses such as image annotation, clustering of image search results, inference of image semantic content, etc. It has been used independently or jointly with image visual contents in CBIR systems. Regardless, it is crucial to extract the right amount of information in order to avoid erroneous cases as illustrated in Figure 2. Webpage segmentation is a basic component to achieve this. So, in the experiments, we present an empirical evaluation of our segmentation algorithm for Web images. We compare our method with VIPS [7], a state-of the-art webpage segmentation algorithm.

*5.2.1 Human-labeled Dataset*

The "segmentation task" described in Section 4.2 provides a human-labeled dataset comprising of 1019 image segments, out of which 898 images were associated to their relevant contextual information. The remaining 121 image segments were header, footer or navigation segments. When identified segments differ between subjects, we chose to consider the bigger section rather than the section that has been defined by at least two subjects. By accepting the bigger sections, we do not lose out on the general topic header relevant to the smaller sections.

In addition, the size of a valid image can be clearly defined from this user study. Most works discarded images with both width and height less than 60 pixels and width-height ratio less than 1/5 or greater than 5 [6, 13, 17]. In the study, users, on top of that, identified images with both height and width less than 60 pixels but greater than 45 pixels. These images are square or rectangular in shape, i.e. with a width-height ratio between ½ and 2. Hence, in addition to the valid image size defined in the literature, we will also extract segments containing square/rectangular images with width and height between 45 and 60 pixels.

*5.2.2 System-based Evaluation*

We evaluate our segmentation algorithm within a system-based framework where the Precision and Recall indicators are used. Precision is the percentage of correctly extracted segments over the total extracted segments and Recall is the percentage of correctly extracted segments over the total actual number of image segments in the dataset. We define *actual* as the images with the expected relevant contextual information; *extracted* as the images and their contextual information extracted by the algorithm; and lastly, *correct* as the extracted images and their contextual information that match the expected ones in *actual*.

The webpages are parsed using the HTMLParser available at http://htmlparser.sourceforge.net/ to obtain their DOM Tree. Our segmentation algorithm is then performed on the resulting DOM Trees. A total of 1012 image segments are *extracted*, slightly less than the *actual* 1019 segments. 748 segments are *correct*, thus, achieving **73% and 72%** for **precision and recall** respectively. The average time taken to process a webpage, extract the images and their contextual information, is ***0.4s***, evaluated on a hardware platform with Duo Core 1.7GHz Intel Pentium Processer and 1GB RAM.

We compare our method to the Vision-based Page Segmentation (VIPS) algorithm, a heuristic DOM-based segmentation algorithm with additional visual information such as horizontal and vertical separators obtained from rendering the webpage. Each resulting visual segment has a defined Degree of Coherence to measure the content consistency within the block, ranging from 1 to 10. The greater the value, the more consistent the content is within the segment. An adjustable pre-defined Permitted Degree of Coherence (PDoC) value is provided to achieve different granularities of content structure to cater for different applications [7]. VIPS is selected as its executable is available at http://www.cs.uiuc.edu/homes/dengcai2/vips/vips.html and it is widely used in many web-based image retrieval systems to extract the image contextual information [6, 17, 25, 39]. For our comparison purposes, we emulate He *et al.*'s work in applying VIPS to extract images and their surrounding texts whereby the PDoC value is empirically set to 5 [17]. Several pages could not be segmented by VIPS due to scripting errors and are therefore excluded from the evaluation. The result is proposed in Table 8.

**Table 8: Performance Comparison using VIPS and our proposed method**

|           | Our Method | VIPS=5 | VIPS=6 | VIPS=7 |
|-----------|------------|--------|--------|--------|
| Actual    | 869        | 869    | 869    | 869    |
| Extracted | 864        | 853    | 853    | 853    |
| Correct   | 628        | 174    | 278    | 333    |
| **Recall**    | **0.72**   | **0.20**   | **0.32**   | **0.38**   |
| **Precision** | **0.73**   | **0.20**   | **0.33**   | **0.39**   |

The table clearly illustrates that our method outperformed VIPS in extracting image segment across a diverse assortment of webpages, mainly because the PDoC value of 5 is not the most optimal value for VIPS to extract image segments. It should be noted that the image segments are considered correct only if the right amount of image contextual information is extracted, no more and no less. VIPS performed poorly because at a PDoC set to 5, it tends to take the bigger section as an image segment. Referring back to Fig 1a, segments 1-4 are considered as one image segment and segments 5-8 as another image segment. If the bigger segments are considered *correct*, then both the algorithms would have achieved over 90% precision. However, bigger sections usually contain lots of noisy information meaningless to the image. Hence, we further test VIPS for higher PDoC values, as shown in Table 8. We stop at a PdoC value equal to 7 as many pages are over-segmented for higher values. This obviously brings about issues for image segmentation especially for webpages with multiple arrangements of images or across a range of diverse contents.

This initial report of performance studies verified the effectiveness of our segmentation algorithm. We believe the precision of the algorithm can be further increased if it were able to tackle deeply structured DOM Trees as in blogs which have a list of posts containing images. Such images are listed images, however, the repetitive sub-tree patterns are found beyond the two changes in the number of texts.

## 6. CONCLUSION

The aim of this paper is to provide a clear definition of the semantically relevant contextual information of Web images and to analyze its characteristics. This is accomplished by a personal observation on 386 Web images, which resulted with a set of hypotheses. A user study is carried out to validate the hypotheses on a dataset of 898 Web images, which allowed us to obtain results of statistical significance.

The first important outcome is the characterization of semantically relevant contextual information with regards to Web images in terms of HTML tag location and semantic nature. This characterization will prove useful in filtering out irrelevant contextual information from relevant ones. We anticipate that the fusion of indexes from image contextual information and indexes derived from low-level content-based characterization will be simpler with the line drawn between tangible (*i.e.* signal, object, and scene) and intangible concepts (*i.e.* abstract) in our proposed semantic concept type characterization. These five classes of semantic concepts can be considered to be at the the highest level of a hierarchy-based classification. Each class will naturally consist of sub-classes, for example the signal concept class can be further divided into sub-classes of color, shape and texture and the scene concept class further partitioned into location, event, collection/aggregation and so forth. Future works involve highlighting this hierarchy of concept classes. Other perspectives pertaining to the analysis of the user study data is related to the study of the types of concepts users frequently suggest in their free description of the image. This would allow us to further refine our characterization of semantically relevant contextual image information.

Another important outcome of these studies verifies that contextual information of any Web image can be defined as a section/segment of texts. We have proposed a novel automatic webpage segmentation algorithm to extract Web images and their semantically relevant contextual information. An experiment validated its effectiveness of 73% precision and 72% recall rates. Our algorithm features DOM Tree-based techniques that do not require any tuning parameters, nor going through underlying browser rendering engines to obtain visual cues. While the approach is highly heuristic, it is however not related to any presentation style or geometry features such as the size, shape or relative position of the image. Hence, we expect our approach to be able to keep up with the inherently dynamic nature of HTML presentation styles. Cascading Style Sheet (CSS) is a style sheet language that is commonly used to define the presentation style of current webpages and is typically stored as a separate document. Our approach is able to handle CSS with just a minor tweak which involves changing the HTML Parser to a parser that is able to render the CSS together with the HTML document into its DOM Tree. Then, our segmentation is able to extract the image segments correctly for most pages using the same heuristics. Its portability to other datasets, however, remains to be tested. It is hoped that extensive testing will be carried out in the future.